\documentclass[reqno,12pt]{article}
\usepackage{amsmath,amsfonts,amssymb,amsthm,amstext,amscd,eucal}
\usepackage{hyperref}
\usepackage{enumerate}
\makeatletter \@addtoreset{equation}{section}

\makeatletter\renewcommand\section{\@startsection {section}{1}{\z@}%
                                   {-3.5ex \@plus -1ex \@minus -.2ex}
                                   {2.3ex \@plus.2ex}%
                                   {\normalfont\large\bfseries}}
\renewcommand\subsection{\@startsection{subsection}{2}{\z@}%
                                     {-3.25ex\@plus -1ex \@minus -.2ex}%
                                     {1.5ex \@plus .2ex}%
                                     {\normalfont\bfseries}}

 \newcommand{\be}{\begin{equation}}
 \newcommand{\ee}{\end{equation}}
 \newcommand{\bea}{\begin{eqnarray}}
 \newcommand{\eea}{\end{eqnarray}}
 \newcommand{\nn}{\nonumber}
 
 \newcommand{\ct}{{\cal T}}

\newcommand{\bse}{\begin{subequations}}
\newcommand{\ese}{\end{subequations}}

\begin{document}
\title{Gravity/CFT correspondence for three dimensional Einstein gravity with a conformal scalar field}
\author{M. Hasanpour\thanks{e-mail:
hasanpour.masoud@gmail.com}, F. Loran\thanks{e-mail:
loran@cc.iut.ac.ir} and H. Razaghian\thanks{e-mail:
razaghian.hamid@gmail.com}\\ \\
  {\it Department of  Physics, Isfahan University of Technology,}\\
{\it Isfahan, 84156-83111, Iran}}
\date{}
  \maketitle
 \begin{abstract}
 We study the three dimensional Einstein gravity conformally coupled to a scalar field. Solutions of this theory are
 geometries with vanishing scalar curvature. We consider solutions with a constant scalar field
 which corresponds to an infinite Newton's constant.  There is a class of solutions with possible
 curvature singularities which asymptotic symmetries are given
 by two copies of  the Virasoro algebra. We argue that the central charge of the corresponding CFT is infinite.
 Furthermore, we construct a family of Schwarzschild  solutions which can be conformally mapped to
 the Mart\'{\i}nez-Zanelli solution of Einstein's equations with a negative
 cosmological constant coupled to conformal scalar field.
 \end{abstract}

\newpage
\tableofcontents


 \section{Introduction}\label{int}
 Asymptotic symmetries of the BTZ black holes \cite{BTZ1,BTZ2} which are solutions of Einstein gravity
 with a negative cosmological constant $\Lambda=-l^{-2}$
 are given by two copies of the Virasoro algebra. The central charge of the corresponding CFT is given by
 the Brown-Henneaux formula \cite{Brown-Henneaux},
 \be
 c=\frac{3l}{2G}
 \ee
 where $l$ is radius of the AdS space and $G$ is  Newton's constant in three dimensions.
 The CFT is believed to be the dual picture
 of the corresponding quantum gravity \cite{ads/cft}. One reason in support of the duality is
 the Strominger's observation \cite{Strominger-97}
 who showed that the Bekenstein-Hawking entropy is
 correctly given by the Cardy formula \cite{Cardy}. In principle, the dual CFT is
 supposed to describe the physics of the gravity side, including the scattering
 processes, Hawking radiation, back reactions, and specially the problem of closed timelike curves in the BTZ solutions and orbifold (orientifold)
 singularities \cite{ortiz-lifschytz, probing-inside, Vijay, Krishnan, BDSS}.

 In this paper, we study the 3D Einstein gravity conformally coupled to a scalar
 field $\psi$. This action gives the bosonic matter term in supergravity coupled to M2-branes
  \cite{Nilsson1, Nilsson2, Nilsson3}. Classical solutions of this theory are geometries
 with vanishing scalar curvature. In this theory Newton's
 constant is scaled by the factor $(1- \pi G\psi^2)^{-1}$. Thus for $\psi^2= 1/\pi G$
  Newton's constant is effectively infinite.  We show that there is a class of  black solutions
 that similar to AdS$_3$ end on a cylindrical  conformal boundary, and the asymptotic symmetries of such geometries are given  by two copies of  the Virasoro algebra.

 Motivated by this observation, we try to extend AdS/CFT correspondence to these asymptotically flat geometries.
 There are  two main difficulties in this problem. First of all, the black hole solutions emerge at
 the critical point where the Planck mass is effectively vanishing. Thus in this case, Einstein gravity
 cannot be considered as a viable effective theory of gravity, since for vanishing  Planck mass, all gravitational
 excitations, whatever they are, become massless at least at tree level.
 From a practical point of view, the first step to identify the dual CFT is to find a reliable
 extension of the existing methods in the literature, to identify the gravitational Noether charges
 corresponding to asymptotic symmetries. In this paper, we argue that similar to AdS/CFT correspondence
 \cite{Bala}, one can still identify the expectation value of the boundary stress tensor with the Brown-York
 tensor \cite{Brown-York} as it  encapsulates the asymptotic geometry, and use the mass scale defined
 by $\psi^2=1/\pi G$ instead of the vanishing Planck mass given by the effectively infinite Newton's constant.

 The second problem is as follows: the Virasoro generators generically scale the size of the conformal boundary.
 We argue that the  only consistent value for the central charge of the boundary CFT is consequently infinity.
 We compute this central charge by three methods; by studying the transformation properties of the Brown-York tensor
  and   computing the Schwartzian  derivative, by computing the Weyl anomaly, and by using the formula given in \cite{Barnich-Brandt,Barnich-Compere}. The central charge appears to be
  \be
  c=\frac{3{\cal R}}{2G}
  \ee
  where $\cal R$ is the radius of the boundary which, unlike the AdS geometry, is infinite in this case.
  We show that if the weights of the CFT states corresponding to black hole solutions are assumed to be given by the Noether charges corresponding to asymptotic symmetries, then the Cardy formula gives a finite entropy. Of course as we discuss in section \ref{sec-entropy} it is questionable to identify this entropy with the Wald's entropy.

 These observations provide new problems in 3D gravity/CFT$_2$ correspondence and the present work  should be considered as a first step towards understanding a corner of it i.e. the gravity/CFT correspondence for Einstein gravity with conformally coupled matter field at the critical point.

  Three primary problems that come to mind are the following. First of all, although in AdS/CFT correspondence it is known that the AdS radius $l\gg l_{\rm Planck}$ and thus $c\gg 1$, such CFTs are not thoroughly studied  in comparison with the ordinary $c\sim{\mathcal O}(1)$
 CFTs. Now we have the signature of a CFT with an infinite central charge. As we review in section \ref{sec-tensionless}, tensionless string theory \cite{WZW} is an example of such a CFT.  Thus from the gravity/CFT correspondence point of view, one  seeks  a general description of CFT's partition function as a function of $1/c$.
 The second problem is to understand the critical point $\psi^2=1/\pi G$ in the frame work of
 \cite{Nilsson1, Nilsson2,Nilsson3} which could also lead to a  realization of the CFT in terms
 of M2-brane configurations. Finally, the black hole solutions of  section \ref{two} possess  curvature singularity located behind the event horizon. Existence of the curvature singularity makes these geometries essentially different
 from the locally AdS (i.e. BTZ) geometries  and their flat limits studied for example in \cite{flat-1,flat-2}. Thus one needs to address the curvature singularities in terms of the boundary CFT.

 The second class of solutions that we study in this paper are the Schwarzschild solutions.
 These solutions can be conformally mapped
 to the Mart\'{\i}nez-Zanelli solutions \cite{MZ} of Einstein's equations with a negative cosmological constant coupled to a conformal scalar field. AdS/CFT
 correspondence is studied for the Mart\'{\i}nez-Zanelli solution \cite{Natsuume,Park}, which has also a curvature singularity behind a horizon. The fact that this
 solution can be conformally mapped to the Schwarzschild solution is a strong evidence for the necessity of extending the AdS/CFT correspondence
 to Einstein gravity without a negative cosmological
 constant. The first example of a such a theory is a CFT that gives the string spectrum in flat space \cite{BMN}.

 The organization of this paper is as follows. In section \ref{two}, we review the Einstein gravity in 2+1 dimensions conformally coupled to a scalar
  field $\psi$ and study the stationary static solutions for $\psi=1/\sqrt{\pi G}$. In section \ref{three} asymptotic symmetries of a class of solutions
  with a cylindrical conformal boundary are studied. In section \ref{four} we study the Schwarzschild and the Mart\'{\i}nez-Zanelli solutions.
  Our results are summarized in section \ref{five}. General stationary static solutions of the theory are given in Appendix \ref{ApA} where we show that
  all stationary static solutions with an $r$-dependent matter field $\psi(r)$ asymptote to $\psi= \sqrt{1/\pi G}$. Gravitational Noether charges and  the corresponding  central charge are computed in appendix \ref{BaBr-appendix}.
  Stability of the Schwarzschild solution against linear perturbations is studied in appendix \ref{ApB}.


 \section{The Model}\label{two}
  Einstein gravity in 2+1 dimensions conformally coupled to a scalar
  field $\psi$ is given by the action
  \be
  \int d^3x
  \sqrt{-g}\left(\frac{R}{2\kappa}-\frac{1}{2}(\partial\psi)^2-\frac{1}{16}R\psi^2\right),
  \label{action}
  \ee
  where $\kappa=8\pi G$.
  Conformal coupling refers to the fact that the matter term in the action is invariant under conformal transformations,
  \be
  g_{\mu\nu}\to \omega^2g_{\mu\nu},\hspace{1cm}\psi\to\psi/\sqrt{\omega}.
  \label{conf-trans}
  \ee
 Under this transformation,
 \be
 R\to
 \omega^{-2}\left[R-4\nabla^2\ln\omega-2(\partial\ln\omega)^2\right].
 \label{conf-R}
 \ee
  where $\nabla$ stands for covariant derivative.

  Adding a cosmological constant ($\Lambda$) term to the action \eqref{action},
  the Einstein field equations become
 \be
 G_{\mu\nu}+\Lambda{g_{\mu\nu}}=\kappa T_{\mu\nu}.
 \label{field-equation}
 \ee
 where $G_{\mu\nu}=R_{\mu\nu}-\frac{1}{2}Rg_{\mu\nu}$ is the Einstein tensor and
 \be
 T_{\mu\nu}=\frac{3}{4}\nabla_{\mu}\psi\nabla_{\nu}\psi-\frac{1}{4}g_{\mu\nu}
    (\nabla\psi)^{2}-\frac{1}{4}\psi\nabla_{\mu}\nabla_{\nu}\psi
    +\frac{1}{4} g_{\mu\nu}\psi\nabla^{\rho}\nabla_{\rho}\psi
    +\frac{1}{8}\psi^{2}G_{\mu\nu}.
 \label{EM tensor}
 \ee
 is the energy momentum tensor.
  One can show that in general, $R=6\Lambda$ on-shell.  This follows from the
 fact that by the conformal symmetry of the matter term, the
 corresponding energy momentum tensor, given in
 Eq.\eqref{EM tensor} is traceless  on-shell,
 \be
 T^{\mu}_{\mu}=\frac{1}{2}\psi\left(\Box-\frac{R}{8}\right)\psi=0
 \ee
 Consequently any solution of the action \eqref{action}, for which $R=0$, can be
 conformally mapped to a solution of the same theory with
 a  cosmological constant $\Lambda$,  if
 \be
 \omega^{-2}\left[-4\nabla^2\ln\omega-2(\partial\ln\omega)^2\right]=6\Lambda.
 \ee
 An example is given  in section \ref{four}.

  Motivated by this observation, we study, in the following, axisymmetric stationary static
  solutions of action \eqref{action}. General axisymmetric stationary static solution with an $r$-dependent
  matter field $\psi(r)$ is discussed in Appendix \ref{ApA}. In the following we concentrate on solutions with
  $\psi=\pm\sqrt{8/\kappa}$. Field equations \eqref{field-equation} with
  $\Lambda=0$ corresponding to action \eqref{action},
    \be
  \left(1-\frac{\kappa\,\psi^2}{8}\right)G_{\mu\nu}=\kappa\,{\tilde T}_{\mu\nu}
  \label{Estn1}
  \ee
  where
  \bea
  \tilde T_{\mu\nu}&=&T_{\mu\nu}-\frac{\psi^2}{8}G_{\mu\nu}\nn\\&=&\frac{3}{4}\nabla_{\mu}\psi\nabla_{\nu}\psi-\frac{1}{4}g_{\mu\nu}
    (\nabla\psi)^{2}-\frac{1}{4}\psi\nabla_{\mu}\nabla_{\nu}\psi
    +\frac{1}{4} g_{\mu\nu}\psi\nabla^{\rho}\nabla_{\rho}\psi.
  \eea
   become trivial in this case and the only condition on the metric comes from  the field  equation for $\psi$
  \be
  \left(\Box -\frac{R}{8}\right)\psi=0,
  \label{psi eom}
  \ee
 which for  a nonvanishing constant $\psi$ implies that
 \be
 R=0.
 \ee
 In general, in three dimensions the Riemann tensor can be determined in terms of the
 Ricci tensor and the metric tensor by the following identity \cite{Carlip-Book},
 \be
 R_{\mu\nu\rho\sigma}=R_{\mu\rho}g_{\nu\sigma}+R_{\nu\sigma}g_{\mu\rho}-R_{\nu\rho}g_{\mu\sigma}-R_{\mu\sigma}g_{\nu\rho}-\frac{1}{2}
 R(g_{\mu\rho}g_{\nu\sigma}-g_{\nu\rho}g_{\mu\sigma}).
 \ee
 Given that $R=0$ there is only one other invariant quantity is the
 Kretschmann scalar,
 \be
 K=R_{\mu\nu\rho\sigma}R^{\mu\nu\rho\sigma}=4 R^{\mu\nu}R_{\mu\nu}.
 \ee
 \subsection{Axisymmetric Stationary Static Solutions}\label{sec-two-one}
 The ansatz for stationary static solutions in $(t,r,\phi)$ coordinate system is given by a diagonal
 metric
 \be
  g_{tt}=-f(r),\hspace{5mm}g_{rr}=\frac{1}{n(r)},\hspace{5mm}g_{\phi\phi}=r^2.
  \label{ansatz}
  \ee
 For this ansatz, the equation $R=0$ can be easily solved to obtain
 \be
 n(r)=\exp\left(-\int
 dr\frac{2rf''(r)f(r)-r{f'}^2(r)+2f'(r)f(r)}{[rf'(r)+2f(r)]
 f(r)}\right),\hspace{1cm}f'(r)\equiv\frac{d}{dr}f(r).
 \label{n}
 \ee

 It is known that Einstein gravity
 \be
 S=\frac{1}{2\kappa}\int d^3x  {R\sqrt{- g}},
  \ee
  has no Black hole solutions \cite{Deser,Ida} while Einstein gravity with a negative cosmological constant enjoys
  the wide class of (orientifolded) BTZ black hole solutions \cite{AyonBeato:2004if, LS1, LS2}. Since all such solutions are locally AdS ,
  they do not possess a curvature singularity which is a common phenomenon in higher dimensional black objects. On the contrary, there are many black
  objects there in Eq.\eqref{n} with curvature singularities.

  In general, a black hole solution is identified  by an event horizon where $n(r)=0$ and
  a region of infinite red-shift where $f(r)=0$. Eq.\eqref{n} simplifies the search for
 possible black hole solutions. An example of such a solution is given by
 \be
 f(r)=r-a,\hspace{1cm}n(r)=\frac{r-a}{(r-\frac{2a}{3})^{4/3}},
 \ee
 which for $a>0$ is a black hole with an event horizon located at $r=a$. Since
 \be
 4K=\frac{7r^2-8\,ra+4a^2}{18\,r^2(r-\frac{2a}{3})^{14/3}},
 \ee
 the curvature singularities at $r=0$ and $r=\frac{2a}{3}$ are
 covered by  the event horizon. For $a<0$ there is a naked curvature singularity at $r=0$.

 As the second example consider the geometry
 \be
 f(r)=r^2-2a^2,\hspace{1cm}n(r)=\frac{r^2-2a^2}{\left|r^2-a^2\right|^{3/2}}.
 \label{the-geometry}
 \ee
 For this solution
 \be
 4K=\frac{3}{2}\frac{r^4-2r^2a^2+4a^4}{\left|r^2-a^2\right|^{5}}.
 \ee
 Thus the curvature singularity at $r^2=a^2$ is behind the event horizon at
 $r^2=2a^2$. This solution can be extended from $r\in{\mathbb  R}^+$ to the $r\in{\mathbb  R}$ region.

 In the next section we discuss gravity/CFT correspondence for this background. Thus it is worth mentioning that similar to BTZ black holes,  the region behind the curvature singularity at $r^2=a^2$ can be removed by {\em folding} the geometry  right there \cite{LS1,LS2}.
 The corresponding  folded geometry is given by the following metric,
 \be
 ds^2=\left[\rho^2\theta(\Phi)+r^2\theta(-\Phi)\right] dt^2+dR^2+\left[\rho^2\theta(-\Phi)+r^2\theta(\Phi)\right]d\phi^2.
 \label{O solution}
 \ee
 where $\Phi=r^2-a^2$,
 \be
 dR^2=-\frac{\left|r^2-a^2\right|^{3/2}}{\rho^2}dr^2=-\frac{\left|\rho^2-a^2\right|^{3/2}}{r^2}d\rho^2,\hspace{1cm}\rho^2=2a^2-r^2,
 \ee
 and $\theta(x)$ is a step function.\footnote{ $\theta(x)$ is zero for $x < 0$, 1 for $x > 0$ and $1/2$ for $x = 0$.}
 This ${\mathbb Z}_2$ folding  is accompanied by insertion of a $\delta$-function source at $\Phi=0$ \cite{Israel}, since there is a
 jump in the curvature  given by \cite{Mansouri-Khorrami},
 \be
 \breve{R}_{\mu\nu}=\left(\frac{1}{ 2 g}[\partial_\mu g]\partial_\nu\Phi-[\Gamma^\rho_{\mu\nu}]\partial_\rho\Phi\right)\delta(\Phi)=
 4a^2\,(-1,0,1)\frac{\delta(\Phi)}{\left|\Phi\right|^{3/2}}.
 \label{KM jump}
 \ee
 Here, $[\Gamma^\rho_{\mu\nu}]$ denotes the jump in the Levi-Civita connections, and $g$ is the determinant of the metric.
 \section{Gravity/CFT duality}\label{three}
 The second example studied in section \ref{sec-two-one} is a special member of an infinite class of solutions identified by the asymptotic
 geometry,
 \be
 ds^2=r^2(-dt^2+d\phi^2)+r\,dr^2.
 \label{thegeometry}
 \ee
 This geometry is, by itself, a solution corresponding to $a=0$ in
 Eq.\eqref{the-geometry}. Similar to AdS$_3$, this geometry ends on a conformal boundary, the
 $(t,\phi)$  cylinder. For AdS$_3$, it known that the asymptotic
 symmetry are given by two copies of Virasoro algebra with a
 central charge proportional to  the AdS$_3$ radius
 \cite{Brown-Henneaux}, which is conjecturally related to a CFT on the
 cylinder. In the following we examine the asymptotic symmetries of the geometry
 \eqref{thegeometry}.
 \subsection{Asymptotic symmetry}
 The asymptotic symmetry group of \eqref{thegeometry} is given by two copies of Virasoro
 algebra.\footnote{For a fixed value of $\psi=\sqrt{8/\kappa}$, one can assume the straightforward boundary condition $\delta\psi=0$
 and reduce the asymptotic symmetries of  the solution to those of the asymptotic geometry. But since the asymptotic geometry \eqref{thegeometry} gives
 $\psi=\sqrt{8/\kappa}+{\mathcal O}(r^{-1/2})$, one can, in principle, extend the asymptotic symmetries to include variation of the scalar field
 $\delta\psi\sim{\mathcal O}(r^{-1/2})$ which is consistent with  \eqref{xi}.}
  To see this in the Brown-Henneaux approach \cite{Brown-Henneaux},
 one needs to change the coordinate system.
 Consider for example, the new radial coordinate $x=r^6$, in terms of which, the geometry \eqref{thegeometry} becomes
 \be
 g_{\mu\nu}= \left(\begin{array}{ccc}
 -x^{1/3}&0&0\\
 &x^{-3/2}&0\\
 &&x^{1/3}
 \end{array}\right)
 \label{first-x}
 \ee
 and assume the following boundary condition for the fluctuations around the geometry,
 \be
 h_{\mu\nu}\sim{\mathcal O} \left(\begin{array}{ccc}
 x^{-2/3}&x^{-1/2}&x^{-2/3}\\
 &x^{-3/2}&x^{-1/2}\\
 &&x^{-2/3}
 \end{array}\right).
 \label{boundary-condition}
 \ee
 The general diffeomorphism preserving the boundary condition \eqref{boundary-condition} is given by
 \be
 \xi=\left(\varepsilon(t,\phi)+{\mathcal O}(\frac{1}{x})\right)\partial_t+\left(\lambda(t,\phi)+{\mathcal O}(\frac{1}{x})\right)\partial_\phi+
 \Big(\alpha(t,\phi) x+{\mathcal O}(1)\Big)\partial_x
 \label{xi}
 \ee
 where
 \be
 \partial_t\varepsilon=-\frac{\alpha}{6}=\partial_\phi\lambda,\hspace{1cm} \partial_\phi\varepsilon=\partial_t\lambda.
  \ee
 Thus the generators of the asymptotic symmetry group are,
 \be
 \xi_n^\pm(\sigma^\pm)=-e^{-in\sigma^\pm}\left(\partial_\pm+3inx\partial_x\right),\hspace{1cm}[\xi_m,\xi_n]_{\rm{Lie}}=-i(m-n)\xi_{m+n}
 \label{1stVir}
 \ee
 where $\sigma^{\pm}=t\pm\phi$ and $\partial_\pm=\frac{1}{2}(\partial_t\pm\partial_\phi)$.

 The existence of such a symmetry group is not a surprise since the conformal boundary of the geometry \eqref{thegeometry} is the conformal
 boundary of an  AdS$_3$ space.\footnote{For AdS spacetime the conformal boundary
  and the causal boundary are the same, but this is not the case for the geometry \eqref{thegeometry}.
 If the field theory dual to the
 gravity is a CFT, it is expected to be located on the conformal boundary. But one may argue that the dual field theory
 is to be located on the causal boundary. If this idea is correct, then the field theory dual to Einstein gravity
 conformally coupled to a scalar field would not be necessarily a CFT. We are grateful to
 M. M. Sheikh-Jabbary for sharing this observation with us.}

 To identify the boundary CFT, one needs to determine the corresponding central
 charge.
 In order to do this, we note that deformations by  $h_{rr}\sim {\mathcal O}(r)$
 effectively scale the volume of the $(t,\phi)$ cylinder since $g_{rr}=r$. Classically such
  deformations   are not observable on the conformal boundary.
  But quantum mechanically they effectively scale
  the central charge which is the vacuum energy in units of the {\em volume} of the $(t,\phi)$ torus.
 To have a meaningful CFT the cental charge should be invariant under such scalings. Thus it is either vanishing or infinite. $c=0$ is outside the domain of gravity/CFT correspondence.\footnote{At least, the Brown-Henneaux formula shows that $c\to0$ corresponds to probing the spacetime in length scales $l\ll l_{\rm Planck}$.} But the $c\to\infty$ case can be understood from the gravity side.
 First of all, in AdS/CFT
 correspondence the central charge is given by the Brown-Henneaux formula
 \be
 c=\frac{3l}{2G}.
 \label{anomaly}
 \ee
 where $R=-6l^{-2}$. Thus the $R=0$ solution correspond to
 $c\to\infty$.\footnote{Although $\psi=\sqrt{8/\kappa}$
 corresponds to $l^{-1}_{\rm Planck}=0$ but since  the energy momentum tensor of a conformally coupled scalar
 field \eqref{EM tensor}
 is traceless by construction, for all solutions of Einstein gravity conformally coupled to a scalar field, $R=0$
 even if  $\psi$ corresponds to a finite Planck length. Thus in \eqref{anomaly} $c^{-1}=0$ for $l^{-1}_{\rm Planck}\to 0$.}
 Furthermore in a
 CFT \cite{Ginsparg}
 \be
 ({L_0})_{\rm cyl}=L_0-\frac{c}{24}.
 \ee
 As  $c\to\infty$, all finite excitations  on the cylinder become effectively degenerate.
 This is in agreement with the fact that for the Einstein gravity conformally
 coupled to  the scalar $\psi=\sqrt{8/\kappa}$, the action is vanishing on-shell, and consequently all solutions
  have equal contribution to the  partition function.

 A dual CFT with  $c\to\infty$ can also be understood in the following way. By AdS/CFT correspondence,
 we know that  a CFT$_2$ with a finite central charge has a dual
 gravity picture in an AdS$_3$ space with radius $l_c=(2G/3)c$. This implies that
 if Einstein gravity with a conformal matter field is dual to a CFT
 with the central charge $c$, then it is also dual to Einstein gravity with the cosmological constant
 $\Lambda_c=-l_c^{-2}$. Such a duality is not reasonable while there is no length scale other than $G$
 in Einstein gravity with a conformal matter field.

   Although it sounds meaningful but it is not a `proof' yet.
  To determine the central charge one needs to compute the corresponding anomaly explicitly. One also
  needs to identify the geometry dual to the vacuum state which energy is $-c/24$.

 In AdS/CFT correspondence, the primary fields of the CFT correspond to the family of the  locally AdS
  solutions including the (orientifolded) AdS space,
 (orientifolded) BTZ and (orientifolded) self-dual orbifolds \cite{Coussaert, LS2}. In principle the CFT will tell us about the BTZ
 singularities, the closed time-like curves and the $\delta$-functions sources of the orientifolded solutions. If the picture obtained so far
 is correct, the $c\to\infty$ CFT would account for the whole family of solutions with $R=0$ that asymptote
 to the geometry \eqref{thegeometry}, including the folded geometry \eqref{O solution} and its odd $\delta$-function source \eqref{KM jump}.

  As the final comment we recall that the $c\to\infty$ CFT is, in principle, {\em included} in any CFT with a finite central charge.
  To see this, recall that for any
  integer $N$, the subalgebra ${\mathcal L}_n\equiv N^{-1}L_{nN}$ of a Virasoro algebra generated by $L_n$
  with central charge $c$, is a Virasoro algebra
  with the central charge $Nc$.\footnote{This orbifolding is studied for example in
  \cite{Borisov:1997nc, Banados:1998wy, Martinec}, and discussed in \cite{LSV3, dBJS}.}
  Obviously the $c\to\infty$ CFT corresponds to {\em orbifolding} a generic CFT by
  $N\to\infty$.\footnote{In AdS/CFT correspondence, $N\to\infty$ is related to the BTZ threshold $M = 0$
  \cite{Martinec}.}

 \subsection{Charges}\label{sec-three-two}
 In this section we compute the central charge of Virasoro algebra \eqref{1stVir},
 and the entropy of the black hole solutions \eqref{the-geometry}.
 For this purpose we first consider a more general asymptotic
 geometry in three dimensions given by the line element
 \be
 {ds^2}=r^2(-d{\ct}^2+d\phi^2\big)+\left(\frac{r}{\ell}\right)^{2z}dr^2\hspace{1cm}z\in{\mathbb R}
 \label{general-asymptotic-geometry}
 \ee
 where
 \be
 \ct=\frac{ t}{\ell},
 \ee
 in which $\ell$ is a parameter characterizing the length scale of the solution. For asymptotically AdS geometry,
 $z=-1$ and $\ell$ is the radius of the AdS space. For
 $z=\frac{1}{2}$ and $\ell=1$, this is the geometry
 \eqref{thegeometry}.

 In general, in order to construct the asymptotic symmetries, one has to define a new radial
 coordinate $x$ by
 \be
 x^b=\left(\frac{r}{\ell}\right),\hspace{1cm}b\in{\mathbb R}^+.
 \ee
 In fact one can show that the asymptotic symmetries are trivial for
 $b(z+1)\ge\frac{1}{2}$.\footnote{This  is implied by  the asymptotic falloff conditions
 $\delta g_{x\ct}\sim \delta g_{x\phi}\sim xg_{xx}\to0$.}
 Therefore for $z\neq -1$ (asymptotic AdS geometry), one assumes that
 \be
 b(z+1)<\frac{1}{2}.
 \ee
 For example, in Eq.\eqref{first-x} where $z=1/2$ we have assumed $b=1/6$.
 Scalar quantities such as the central charge are independent of the
 choice of  radial coordinate, so their values do not depend on  $b$.
 In terms of the new radial coordinate $x$ the asymptotic line
 element is
 \be
ds^2=ds_{\cal B}^2+N^2dx^2
 \ee
 where
 \be
 ds_{\cal B}^2= \gamma_{\mu\nu}d\sigma^\mu d\sigma^\nu,\hspace{1cm}N=b\,\ell\,x^{(zb+b-1)}
 \ee
 and
 \be
 \gamma=r^2{\rm diag}(-1,1)
 \ee
 \subsubsection{Asymptotic symmetry}
 Consider the vector
 \be
 \xi=\big(\epsilon+\frac{\bar\epsilon}{x^{2b}}\big)\partial_\ct+
 \big(\lambda+\frac{\bar\lambda}{ x^{2b}}\big)\partial_\phi+
 \alpha\, x\,\partial_x
 \ee
   and define the asymptotic conformal Killing vector
 \be
 \delta
 g_{\mu\nu}=\partial_\nu\xi^{\alpha}g_{\alpha\mu}+\partial_\mu\xi^{\alpha}g_{\alpha\nu}+\xi^\alpha\partial_\alpha
 g_{\alpha\mu}+2\rho g_{\mu\nu},
 \label{tuned}
 \ee
 in which $\rho=-b(z+1)\alpha$ such that $\delta
 g_{xx}=0$.\footnote{This fall off condition on $g_{xx}$ is not crucial in what follows and instead of the asymptotic conformal Killing vectors, one can define the asymptotic symmetry by a generalization of  Eq.\eqref{boundary-condition}, for which $\delta g_{xx}\sim g_{xx}$. The corresponding vector $\xi$ is given by
 Eq.\eqref{BaBr-Killing}.  The essential data is Eq.\eqref{deformed-metric} which is independent of this choice.}
   One can verify that for
  \be
  \begin{array}{ccc}
 \dot\epsilon+b\,\alpha+\rho=0,&&2\,\bar\epsilon-bz\,\dot\alpha=0,\\
 \lambda'+b\,\alpha+\rho=0,&&-2\,\bar\lambda-bz\,\alpha'=0,\\
 \dot\lambda=\epsilon',&&
 \end{array}
 \label{guess}
 \ee
 where for example, $ \alpha'=\partial_\phi\alpha,$ and
 $\dot\alpha=\partial_\ct\alpha$,
  \be
 \xi=\xi(x^\pm)=\sum_{m\in{\mathbb
 Z}}\xi^\pm_m\exp[im(x^\pm)],\hspace{1cm}x^\pm=\ct\pm\phi.
 \ee
 Furthermore,
 \bea
 i[\xi^+_m,\xi^+_n]_{\rm Lie}&=&(m-n)\xi^+_{m+n},\nn\\
 i[\xi^-_m,\xi^-_n]_{\rm Lie}&=&(m-n)\xi^-_{m+n},\nn\\
 {[\xi^+_m,\xi^-_n]}_{\rm Lie}&=&0.
  \label{virasoro}
  \eea
 and  $\xi$ generates the following diffeomorphism
 \be
 \gamma_{\mu\nu}=r^2\gamma^{(0)}_{\mu\nu}\to\gamma_{\mu\nu}=r^2\gamma^{(0)}_{\mu\nu}+\ell^2\,\gamma^{(1)}_{\mu\nu}+\cdots,
 \ee
 in which
 \be
 \gamma^{(0)}={\rm diag}
 (-1,1),\hspace{1cm}\gamma^{(1)}_{\pm\pm}=-\frac{\partial^3\epsilon^\pm}{\partial{x^{\pm}}^3},\hspace{1cm}\gamma^{(1)}_{+-}=0.
 \label{deformed-metric}
 \ee
 In the following we compute the corresponding charges and the central charge. Similar results are obtained in Appendix \ref{BaBr-appendix} using the formula given in \cite{Barnich-Brandt}.
  \subsubsection{Boundary stress tensor}
 Consider the Brown-York stress tensor \cite{Brown-York} defined by
 \be
  \tau_{\mu\nu}=\frac{1}{8\pi
 G}(K_{\mu\nu}-K\gamma_{\mu\nu})
 \ee
 where $K_{\mu\nu}$ is the extrinsic curvature of the boundary defined by
 \be
 K_{\mu\nu}=-\gamma_\mu^\alpha\nabla_\alpha n_\nu.
 \ee
  $n^\mu$ is the outward pointing unit vector to the
 boundary, and the boundary metric $\gamma_{\mu\nu}$ is defined
 by the ADM-like decomposition of the metric
 \be
 ds^2=N^2dr^2+\gamma_{\mu\nu}(dx^\mu+N^\mu dr)(dx^\nu+N^\nu dr).
 \ee
  For Einstein gravity minimally coupled to matter fields, one can
  show that after  subtracting the `vacuum' contribution to
  $\tau$,
  \be
  {\cal D}_\mu\tau^{\mu\nu}=-T^{\alpha\beta}n_\beta\gamma^\nu_\alpha
  \ee
 where  ${\cal D}_\mu$ is the covariant derivative compatible with
 $\gamma_{\mu\nu}$ and $T_{\mu\nu}$ is the matter field stress tensor. Consequently the charges of the spacetime are encoded
  in the Brown-York stress tensor \cite{Brown-York}.

  Inspired by this result and noting that the essence of the Brown-York tensor is the geometry, we postulate that for any spacetime
  \eqref{general-asymptotic-geometry}, the charges corresponding to
  the asymptotic symmetries \eqref{virasoro}   are given by  the regularized tensor\footnote{In general the coefficient $(8\pi G)^{-1}$
 is  given by the gravitational mass scale of the theory. For example for BTZ black holes, $G$ is the  Newton's constant in three dimensions.
 For the black hole solution \eqref{the-geometry} the Planck mass is effectively zero but there is another gravitational mass scale given by
 the conformal scalar field $\psi^2=(\pi\,G)^{-1}$.}
 \be
  \tau^{\rm{reg}}_{\mu\nu}=\frac{1}{ 8\pi
 G}\left(K_{\mu\nu}-\bar K\gamma_{\mu\nu}\right),\hspace{1cm}\bar
 K=K-\frac{K^{(0)}}{2}
  \label{def-of-reg}
  \ee
 where $K^{(0)}_{\mu\nu}$ is the extrinsic curvature of the
 `vacuum' solution $\gamma_{\mu\nu}=r^2\gamma^{(0)}_{\mu\nu}$.\footnote{This regularization is
 consistent with the one introduced in \cite{Bala} for $AdS_3$ where $K^{(0)}=-2\ell^{-1}$.}
 For the asymptotic geometry one can verify that
  \begin{enumerate}
 \item
  $\tau^{\rm reg}_{\mu\nu}$ is a symmetric tensor with respect to
 $\gamma_{\mu\nu}$
\item
  {$\mbox{Tr}\tau^{\rm reg}=0$}
\item
  ${\cal  D}^\mu\tau^{\rm reg}_{\mu\nu}=0$
 \end{enumerate}
 The first identity is proven in \cite{Brown-York}. The second identity is trivial for $K_{\mu\nu}=K^{(0)}_{\mu\nu}$.
 In other cases there is a trace anomaly which is related to the central charge of the CFT on the boundary. We discuss it in the
 following. To prove the third identity, note that
 \be
 K_{\mu\nu}=-\frac{\partial_x g_{\mu\nu}}{2\sqrt{g_{xx}}}
 \label{K-simple}
 \ee
 For the metric
 \be
 \gamma_{\mu\nu}=r^2\gamma^{(0)}_{\mu\nu}+\ell^2\gamma^{(1)}_{\mu\nu}+\cdots
 \label{metric-again}
 \ee
 Eq.\eqref{K-simple} can be used to show that
 \be
 K_{\mu\nu}=-\frac{r\gamma^{(0)}_{\mu\nu}}{\sqrt{g_{rr}}}+\cdots
 \label{Kmunu-again}
 \ee
 and
 \be
 K=-\frac{2}{r\sqrt{g_{rr}}}+\frac{\ell^2{\gamma^{(0)}}^{\mu\nu}\gamma^{(1)}_{\mu\nu}}{r^3\sqrt{g_{rr}}}+\cdots.
 \label{K-again}
 \ee
 Thus,
 \be
 \tau^{\rm reg}_{\mu\nu}=\frac{1}{8\pi
 G}\frac{\ell^2}{\cal R}\gamma^{(1)}_{\mu\nu}
 \label{mytau}
 \ee
  in which $\cal R$ is the radius of the boundary. By definition
 \be
 K^{(0)}=-\frac{2}{\cal R}.
 \label{radius-of-boundary}
 \ee
 The above mentioned properties of $\tau^{\rm reg}$ follows from
 Eq.\eqref{mytau} and \eqref{deformed-metric}.
 We postulate that $\tau^{\rm reg}_{\mu\nu}$ corresponds to the CFT
 stress  tensor.
 \subsubsection{Central charge}
 In general given a CFT on a plane with metric $ds^2=-dw^+dw^-$ the
 diffeomorphism
 \be
 w^\pm\to w^\pm-\zeta^\pm(w^\pm)
 \ee
 results in
 \be
 T_{\pm\pm}\to T_{\pm\pm}+(2\partial_\pm\zeta^\pm
 T_{\pm\pm}+\zeta^\pm\partial_\pm
 T_{\pm\pm})-\frac{c}{24\pi}\partial_\pm^3\zeta^\pm
 \ee
 where $T_{\mu\nu}$ is the CFT stress tenor.
  Assuming that (the dimensionless coordinates on the boundary) $w^\pm$
  are given by
 \be
 \ell\, w^\pm={\cal R}\,(\ct\pm\phi)
 \label{boundary-coordinates}
 \ee
 then Eq.\eqref{deformed-metric} gives
  \be
  \gamma^{(1)}_{\pm\pm}=-
  \frac{{\cal R}^2}{2\ell^2}\partial^3_\pm\zeta^\pm,\hspace{1cm}
  \zeta^\pm=\frac{2\cal R}{\ell}\epsilon^\pm.
  \ee
 Assuming that $T_{\mu\nu}=\tau^{\rm reg}_{\mu\nu}$ one obtains
 \be
 c=\frac{3}{2}\frac{\cal R}{G}
 \label{central-charge-1}
 \ee
 This result strictly depends on the choice made in (the right column of) \eqref{guess} which in particular works for $z\neq0$. This is not a flaw in the model
 since  uniqueness of the asymptotic symmetry is not claimed so far.
 In the following, we compute the central charge in terms of the trace
 anomaly and obtain the same result for general values of $z$.

  The stress tensor of a CFT$_2$ has a trace anomaly
 \be
 \mbox{Tr}\tau^{\rm reg}=-\frac{c}{24\pi}{^{(2)}\!R}.
 \ee
  To calculate the trace anomaly one can use the identity
 \be
 G_{\mu\nu}n^\mu
 n^\nu=-\frac{1}{2}\big({^{(2)}\!R}+K_{\mu\nu}K^{\mu\nu}-K^2\big)
 \label{GR-identity}
  \ee
  where $G_{\mu\nu}$ is the Einstein tensor.
 For a geometry given by \eqref{deformed-metric},
 \be
 G_{\mu\nu}n^\mu
 n^\nu=\frac{1}{r^2g_{rr}}+ \cdots.
 \ee
 Using Eqs.\eqref{metric-again}, \eqref{Kmunu-again} and \eqref{K-again}, one obtains
 \be
 {^{(2)}\!R}=\frac{2\ell^2{\gamma^{(0)}}^{\mu\nu}\gamma^{(1)}_{\mu\nu}}{r^4{g_{rr}}}=K^{(0)}(K^{(0)}-K).
 \ee
 On the other hand,  Eq.\eqref{def-of-reg} gives
 \be
 \mbox{Tr}\tau^{\rm reg}=\frac{K^{(0)}-K}{8\pi G}.
 \ee
 Consequently,
 \be
 c=\frac{3\cal R}{2G}.
 \ee
 In Eq.\eqref{central-charge-1} the same result was obtained for $z\neq0$. Thus it is legitimate
 to assume the validity of this result for general values of $z$ and classify
 the asymptotic spacetimes with respect to the corresponding central
 charges,
 \be
  c=\frac{3\cal R}{2G}=\lim_{r\to\infty}\frac{3\ell}{2G}\left(\frac{r}{\ell}\right)^{1+z}=\left\{\begin{array}{llll}0&&z<-1\\
 \frac{3\ell}{2G}&&z=-1&AdS_3\\ \infty &&z>-1\end{array}\right.
 \label{final-central-charge}
 \ee
 Recall that this is the central charge of the Virasoro algebra \eqref{virasoro} of gravitational charges
 corresponding to symmetries of the asymptotic geometry
 \eqref{general-asymptotic-geometry}. The black hole solution
 \eqref{the-geometry} corresponds to $z=1/2$ and thus the
 correpsonding central charge is infinite.
 In appendix \ref{BaBr-appendix} we obtain the same value for the central charge
 by using the formula given in \cite{Barnich-Brandt}, where we also
 compute the Noether charges corresponding to symmetries of the asymptotic
 geometry.

 It is worth mentioning that the formula
 \eqref{final-central-charge} for $z\ge -1$ (including the BTZ
 black hole  and the black hole geometry \eqref{the-geometry}
 )  is consistent with the $c$-theorem \cite{c-theorem}. Recall that holography implies that
 an IR cut-off $r_{\rm IR}$  corresponds to a UV cut-off $\Lambda_{\rm UV}=r_{\rm IR}$  on the CFT
 side \cite{Susskind}. At such a cut-off Eq.\eqref{central-charge-1} gives a finite central charge
 \be
 c=\frac{3\ell}{2G}\left(\frac{r_{\rm IR}}{\ell}\right)^{z+1}.
 \ee
 which,  for $z\ge -1$,   is an increasing function of the UV cut-off $\Lambda_{\rm UV}=r_{\rm IR}$  with a fixed point $c\to\infty$ at $\Lambda_{\rm
 UV}\to\infty$. We discuss a related topic at the end of section \ref{sec-entropy}.

 \subsubsection{Mass}
 The mass of a solution can be defined by
 \be
 M=\lim_{r\to\infty}2\pi {\cal R}\tau^{\rm reg}_{tt}
 \label{def-of-mass}
  \ee
 This definition is motivated by following facts:  $\tau^{\rm reg}_{tt}$ is
 the energy density and the `volume' equals $2\pi {\cal R}$. In order to
 show that this definition is the correct one, we compute the mass
 for a geometry given by Eq.\eqref{metric-again} in which
 \be
 \gamma^{(1)}_{\pm\pm}=2GM_0,\hspace{1cm}\gamma^{(1)}_{+-}=0
 \label{massive-geometry}
 \ee
 Using Eq.\eqref{mytau} one verifies that,
 \be
 M=M_0.
 \ee
  It is useful to give the geometry corresponding to \eqref{massive-geometry},
 \be
 ds^2=\left(\frac{r}{\ell}\right)^{2z}dr^2+r^2(-d\ct^2+d\phi^2)+4GM\ell^2(d\ct^2+d\phi^2)
 \ee
 in  the ansatz \eqref{ansatz}. Defining a new radial coordinate
 $\rho^2=r^2+4GM\ell^2$ one verifies that the asymptotic geometry is
 given by
 \be
 ds^2=-(\frac{\rho^2}{\ell^2}-8GM)dt^2+\left(\frac{\rho}{\ell}\right)^{2z}\frac{\rho^2d\rho^2}{\rho^2+4(z-1)GM\ell^2}+\rho^2d\phi^2
 \label{deformed-geometry}
 \ee
 This is a key result.  For $z=-1$ this is the static BTZ and for $z=\frac{1}{2}$ this is
 the asymptotic geometry for the black hole solution
 \eqref{the-geometry}. It is interesting to verify this `holographic prediction' for black hole
 solutions  to other theories of $3D$ gravity with asymptotic
 geometry  \eqref{general-asymptotic-geometry}.

 The geodesic equation for a point particle  initially at rest at radial infinity,
 \be
 \frac{d^2\,\rho}{dt^2}=-\frac{\rho^2+4GM\ell^2(z-1)}{\rho\ell^2}\left(\frac{\ell}{\rho}\right)^{2z}+\cdots.
 \ee
  implies that for $z<1$, the mass term produces a repulsive force.
 \subsubsection{Conformal matter at critical value}\label{sec-tensionless}
 In section \ref{two}, for Einstein gravity conformally coupled to a scalar field at the critical
 value $\psi=(\pi G)^{-1/2}$ we found a black hole solution \eqref{the-geometry} which asymptotic geometry is given by Eq.\eqref{deformed-geometry}
 with $z=\frac{1}{2}$ and the corresponding central charge \eqref{final-central-charge} is
 infinite.  For this solution the Planck  mass is effectively zero
 \be
 M_{\rm Pl}^{\rm eff}=\left(1-\frac{\psi^2}{\pi G}\right)M_{\rm
 Pl}.
 \ee
  In string theory this limit corresponds to $ c\to
 \infty$ \cite{WZW}. In fact, in WZW models,
 \be
 \alpha'=\frac{1}{{k-g^\vee}},\hspace{1cm}\tilde
 c=\frac{{({\rm dim} G)k}}{{k-g^\vee}},
 \ee
 where $k$ is the level of current algebra and $g^\vee$ is the dual Coxeter number of
 $G$. The critical level is given by $k=g^\vee$.
 \subsubsection{Entropy}\label{sec-entropy}
 The black hole solution  \eqref{the-geometry} has a finite mass and
 a finite Hawking temperature,
 \be
 T=(2\pi\ell)^{-1}\sqrt{\frac{2}{ a}}=(2\pi\ell)^{-1}(GM)^{-1/4}.
 \label{temperature}
 \ee
 In Eq.\eqref{the-geometry}, $\ell=1$. In principle, one
 can use the first law of thermodynamics
 \be
 dM=TdS_{\rm c}
 \ee
 to compute the canonical entropy of the black hole,
 \be
 S_{\rm c}=\left(\frac{2}{5}\right)\frac{\cal A}{4G},\hspace{1cm}{\cal A}=2\pi\sqrt{-\det g(r_{\rm h})}.
 \label{area-of-horizon}
 \ee
 This entropy is finite while the  microcanonical entropy given by the Cardy formula
 \be
 S_{\rm mc}=2\pi\sqrt{\frac{c \Delta}{6}}+2\pi\sqrt{\frac{c \bar\Delta}{6}}
 \ee
 where $\Delta=\bar\Delta=\frac{M\ell}{2}$,  is infinite. The point is that the black hole solution \eqref{the-geometry} has a negative heat capacity as can be seen from equation
 \eqref{temperature}. Thus it never comes to equilibrium
 with an infinite heat bath. Thus  canonical entropy is not well
 defined in this case \cite{Wald-living}.

 Before closing this section, we report an observation for which we do not have a clear justification.
 As we show in Eq.\eqref{BaBr-Q},  the Noether charge for symmetries of the  asymptotic geometry
 \eqref{deformed-geometry} is given by
 \be
 Q_\xi=M\,\frac{(1-z)}{2}\frac{\ell}{\cal R}\,\xi^t
 \ee
 Consequently one may assign the following weights to the corresponding CFT state
 \be
 {\bar\Delta}'=\Delta'= \frac{(1-z)}{4}\frac{M\ell^2}{\cal R}
 \ee
 In this case, the  Cardy formula gives a finite entropy
 \be
  S'_{\rm mic}=2\pi\sqrt{\frac{(1-z)}{2}\frac{M\ell^2}{2G}}
 \label{second-Cardy}
 \ee
 which  agrees with the `naive' area law
 \be
 \tilde S=\frac{2\pi r_+}{4G}
 \label{naive}
 \ee
 where $r_+$ is given by $g_{\rho\rho}$ in the asymptotic geometry
 \eqref{deformed-geometry},
 \be
 r_+=4(1-z)\,GM\ell^2
 \ee
 We call  formula \eqref{naive} naive because the area of event horizon is given by
 ${\cal A}$ defined in Eq.\eqref{area-of-horizon} which is not in general equal
 to $2\pi r_+$. It should be noted that  $r_+$ is not  even the actual radius of event horizon unless $z=-1$
 (BTZ geometry). Seemingly, this result implies that  the boundary CFT observes the asymptotic geometry \eqref{deformed-geometry}
 and interprets it as a black hole geometry with an event horizon located at
 $r_+$. Conceptually this is a reasonable statement since
 $r\to\infty$ corresponds to the IR limit on the gravity side of
 gravity/CFT correspondence, and in the IR limit, an observer naturally probes the asymptotic geometry and
 is blind to the details of the spacetime structure.

 \section{2+1 dimensional Schwarzschild solution}\label{four}
 In this section we study the Schwarzschild solution. As we show in the following,
 although this geometry does not belong to the class of solutions studied in the previous section,
 it plays a role in gravity/CFT correspondence since it can be conformally mapped to the Mart\'{\i}nez-Zanelli
 solution \cite{MZ}.

 The Schwarzschild solutions is given by
 \be
 n(r)=f(r)=1+\frac{a}{r}.
 \label{schwarzschild}
  \ee
  The curvature singularity in
  \be
 R^{(2)}=\frac{3}{2}\frac{a^2}{r^6}
 \ee
 is hidden by an event horizon if $a<0$. This condition also
 results in a downward gravitational pull of the black hole.
 \subsection{Rotating Solutions}
 The  {\em rotating} Schwarzschild solution is given by the metric
 \be
 ds^2=-f(r)dt^2+\frac{dr^2}{f(r)}+r^2(d\phi+N^\phi dt)^2,
 \ee
 in which
 \be
 f(r)=\left(1-\frac{2M}{r}+\frac{J^2}{r^2}\right),\hspace{1cm}N^\phi=\frac{J}{r^2}.
 \ee

 The Kerr solution is given by the metric,
 \be
 ds^2=-\frac{\Delta(r)-a^2}{\Sigma(r)}dt^2- 2a\frac{r^2+a^2-\Delta(r)}{\Sigma(r)}dtd\phi+\frac{(r^2+a^2)^2-\Delta(r)a^2}{\Sigma(r)}d\phi^2
 +\frac{\Sigma(r)}{\Delta(r)}dr^2,
\label{Kerr}
\ee
 where, $\Sigma(r)=r^2$ and $\Delta(r)=(1+Q^2)r^2-2Mr+a^2$, and $a=\frac{J}{M}$ in which
 $J$ is the total angular momentum. For $Q=0$ this solution corresponds to the geometry
 at the equator ($\theta=\frac{\pi}{2}$) of the Kerr solution in four dimensions.

 For these solutions,
 \be
 R=0,\hspace{1cm}R^{(2)}=\frac{6M^2}{r^6}.
 \ee

 \subsection{The Mart\'{\i}nez-Zanelli solution}
 The Mart\'{\i}nez-Zanelli solution is a black hole solution of Einstein gravity with a negative cosmological constant
 $-l^{-2}$  conformally coupled to a massless scalar field. The solution is given by the metric
 \be
 ds^2=-F(\rho)dt^2+\frac{d\rho^2}{F(\rho)}+\rho^2d\phi^2,\hspace{1cm}F(\rho)=\frac{(2\rho+\rho_0)^2(\rho-\rho_0)}{4l^2\rho},
 \ee
 and
 \be
 \Psi(\rho)=\sqrt{\frac{8\rho_0}{\kappa(2\rho+\rho_0)}},
 \label{Psi}
 \ee
 where  $\rho_0>0$ denotes the radius of the event horizon.
 To show that this solution can be obtained
 from the Schwarzschild solution \eqref{schwarzschild} by a conformal
 transformation,
 \be
 g_{\mu\nu}\to\omega^2(r)g_{\mu\nu},\hspace{1cm}\psi(r)\to\frac{1}{\sqrt{\omega(r)}}\psi(r).
 \label{conf.trans}
 \ee
  one may rewrite the Schwarzschild solution as
 \be
 ds^2=-(1+\frac{a}{r})\alpha^2 dt^2+\frac{\lambda^2}{1+\frac{a}{r}}dr^2+r^2d\phi^2,
 \ee
 where $\alpha$ and  $\lambda$ are, for the moment, arbitrary constants.
 By performing the above conformal transformation on the Schwarzschild solution one obtains a spacetime for which
 \be
 R=-\frac{2}{\lambda^2 r\omega^4(r)}\left\{(r+a)\left[2\omega(r)\omega''(r)-{\omega'}^2(r)\right]+
 2\omega(r)\omega'(r)\right\}.
 \ee
 Solving this equation for $R=-6/l^2$, a solutions is
 \be
 \omega(r)=\frac{\beta}{r\left(1+\frac{3a}{2r}\right)},\hspace{1cm} \beta=\frac{3a}{2}
 \label{omega},
 \ee
 where $\beta$ is determined by the Einstein field equation
 \be
 G_{\mu\nu}-\frac{g_{\mu\nu}}{l^2}=\kappa T_{\mu\nu}.
 \ee
   $T_{\mu\nu}$ is the matter stress tensor Eq.\eqref{EM tensor},
 and
  \be
 \psi(r)=\sqrt{\frac{8}{\kappa\omega(r)}}.
 \label{psi}
 \ee
 For $\omega(r)$ given by Eq.\eqref{omega}, the Ricci scalar is $R=-8/(9\lambda^2a^2)$ which
 determines  $\lambda^2=4l^2/(27a^2)$.
 Defining $\rho^2\equiv g_{\phi\phi}= r^2\omega^2(r)$, (which directly converts Eq.\eqref{psi} to Eq.\eqref{Psi}),
 the Mart\'{\i}nez-Zanelli metric is obtained for $\alpha=\lambda^{-1}$
 and $\rho_0=- 3a$.

 The conformal map between the Schwarzschild solution and the Mart\'{\i}nez-Zanelli solution is specially useful in studying scattering in
 the Mart\'{\i}nez-Zanelli background. Scattering in the Schwarzschild background is thoroughly studied in the literature. Using the conformal map above, all
 those results can be applied to the Mart\'{\i}nez-Zanelli
 background.

  The relation between the Schwarzschild solution and the Mart\'{\i}nez-Zanelli solution can be analyzed from a different point of view.
 The Schwarzschild spacetime and the Mart\'{\i}nez-Zanelli spacetime are static and stationary, i.e. in both cases \cite{Hortacsu},
  \be
  ds^2=-f(r)dt^2+\frac{dr^2}{f(r)}+r^2d\phi^2.
  \label{des-ansatz}
  \ee
 For gravity conformally coupled to matter field the Einstein equation,
  \be
  G_{\mu\nu}+\Lambda g_{\mu\nu}=\kappa T_{\mu\nu},
  \ee
  indicates that $R=6\Lambda$. Solving this equation for $f(r)$ one obtains $f(r)=-\Lambda r^2+\frac{A}{r}+B$. Solving for the matter field $\psi$,
  one obtains,
  \be
  \psi=\sqrt{\frac{8A}{\kappa(A+\frac{2Br}{3})}},\hspace{1cm}\Lambda=-\frac{4B^3}{27A^2}.
  \label{dis-psi}
  \ee
 It is clear that for $\Lambda<0$ this is the Mart\'{\i}nez-Zanelli solution while for
 $\Lambda=0$, Eq.\eqref{dis-psi} retrieves the Schwarzschild solution
  $\psi=\sqrt{8/\kappa}$.
  The conformal factor that gives the conformal map between these solutions can be easily
  obtained: the equations,
  \bea
  F(\rho)&=&f(r)\omega^2(r),\nn\\
  \frac{d\rho^2}{F(\rho)}&=&\omega^2(r)\frac{dr^2}{f(r)},
  \eea
  give $d\rho=\omega^2(r)dr$ and using the definition $\rho\equiv r\omega(r)$, one obtains the conformal factor given by Eq.\eqref{dis-psi}.
   Stability of the Schwarzschild solution against linear perturbations is studied in appendix \ref{ApB}.
 \section{Summary}\label{five}
 We studied 3D Einstein gravity conformally coupled to a massless
 scalar field $\psi$. Solutions of this theory are
 geometries with vanishing Ricci scalar. We studied stationary
 static solutions with $\psi=\sqrt{8/\kappa}$ including the Schwarzschild solution.
 We explicitly showed that the Schwarzschild solution
 can be conformally mapped to the Mart\'{\i}nez-Zanelli solution,
 and similar to it, the Schwarzschild geometry is
 unstable against linear perturbations.

 Furthermore, we observed that $R=0$ has an infinite class of stationary
 static solutions which similar to AdS$_3$ end on a cylindrical  conformal boundary. Following the Brown-Henneaux approach, we showed that
 the  asymptotic symmetries of these solutions are given by two
 copies of the Virasoro algebra. We argued that the central charge of the dual CFT is infinite.
 In fact, a finite central charge is not consistent with the
 asymptotic symmetries where deformations that scale the
 $g_{rr}$ component of the metric are allowed. Furthermore we argued that an infinite central charge is
 consistent  with considerations concerning the semiclassical partition function. Using three different methods we obtained the following value for the central charge
 \be
 c=\frac{3{\cal R}}{2G},\hspace{1cm} {\cal R}=-\frac{2}{K^{(0)}}
 \ee
 in which $K^{(0)}$ denotes the extrinsic curvature of the boundary.
 \section*{Acknowledgement}
 We would like to thank M. M. Sheikh-Jabbari and H. Soltanpanahi for valuable  comments and discussions.

 \appendix
 \section{General axisymmetric stationary static solutions}\label{ApA}
 In this section we give the most general static stationary solution of Einstein gravity in three dimensions coupled to a conformal scalar
 field, given by action \eqref{action}.   The corresponding  Einstein field equations are
  \be
 G_{\mu\nu}=\kappa T_{\mu\nu},
 \ee
  where $G_{\mu\nu}$ is the Einstein tensor and $T_{\mu\nu}$ is the matter stress tensor given in Eq.\eqref{EM tensor},
  and the matter field equation is given by Eq.\eqref{psi eom}.
 \subsection*{Constant $\psi$}
 In this case either $\psi=\sqrt{8/\kappa}$ and $R=0$ or $\psi\neq\sqrt{8/\kappa}$ and $R_{\mu\nu}=0$. We studied the first case in section \ref{two}.
 The second case is the traditional Einstein equation for vacuum, because as can be readily seen in the action \eqref{action},
 a nonvanishing  constant $\psi$ only changes the Newton's constant.
 \subsection*{$r$-dependent solution}
 It is well worth studying this case since it is essentially an example of Einstein gravity with varying Newton's constant.
 Even for a varying $\psi(r)$, the Ricci scalar is vanishing on-shell because $T^\mu_\mu=0$ by construction. For the metric ansatz,
 \be
 ds^2=-f(r)dt^2+\frac{dr^2}{n(r)}+r^2d\phi^2,
 \ee
 the most general solution of Einstein field equations is
 \bea
 f(r)&=&a\exp\left({\int^r d\rho\frac{\psi'(2\rho\psi'+\psi)}{\frac{2}{\kappa}-\frac{1}{4}(\rho\psi^2)'}}\right),\nn\\
 n(r)&=&b\exp\left({\int^r d\rho\frac{\psi\psi'-\rho{\psi'}^2+\rho\psi\psi''}{\frac{2}{\kappa}-\frac{1}{4}(\rho\psi^2)'}}\right),
 \label{f and h}
 \eea
 where $\psi(r)$ is given by the following equation,
 \be
  \left(1-\sqrt{\frac{\kappa}{8}}\psi\right)^{1+\frac{c}{2}}\left(1+\sqrt{\frac{\kappa}{8}}\psi\right)^{1-\frac{c}{2}}=\frac{d}{r},
 \label{Appsi}
 \ee
 in which $d>0$ and $c$ are arbitrary constants. Since $c\leftrightarrow-c$ corresponds to $\psi\leftrightarrow-\psi$ which is a symmetry of the action,
 one can take $c\in{\mathbb R}^+$. $c=0$ is not
 allowed because for $c=0$, $(r\psi^2)'-\frac{2}{\kappa}=0$ and thus gives a singularity in Eq.\eqref{f and h}. In fact,
 independent of Eqs.\eqref{f and h}, one can show that the only solution of equation $(r\psi^2)'-\frac{2}{\kappa}=0$
 that solves the Einstein field equations is $\psi=\sqrt{8/\kappa}$.

 For $c>0$, Eq.\eqref{Appsi} implies that asymptotically,
 \be
 \lim_{r\to\infty }\psi(r)=\pm\sqrt{\frac{8}{\kappa}}.
 \ee
 The geometry \eqref{thegeometry} corresponds to $c=2$.
 \section{Noether Charges}\label{BaBr-appendix}
 In section \ref{two}, we observed that Einstein-Hilbert action with conformal matter at the critical value $\psi_0=\pm\sqrt{8\kappa^{-1}}$ admits black hole solutions.  So far we have  treated
   \be
   G_{\rm eff}=G(1-\frac{\kappa}{8}\psi^2)^{-1}
   \ee
   as the effective coupling constant, and considered the value $\psi_0$ as the critical point
   where the Planck mass vanishes. In this section, in order to obtain the conserved charges corresponding
   to asymptotic Killing vectors  \cite{Barnich-Brandt,Barnich-Compere} we rewrite the action in the Einstein frame.
   Using the conformal transformation $g\to \exp(2\omega) g$
   where\footnote{$e^{-\omega}$  is well defined provided that the effective Newton's constant is nonnegative
   i.e. $\left|\psi\right|\le\psi_0$.}
 \be
 e^{-\omega}=\left(1-\frac{\kappa}{8}\psi^2\right)
 \label{BaBr1}
 \ee
 one obtains,
 \be
 S=\frac{1}{2\kappa}\int\sqrt{-g}\left[R-4\nabla^2\omega-2\left(1+\frac{e^{-\omega/2}}{2\kappa\sinh\frac{\omega}{2}}\right)
 (\partial\omega)^2\right]
 \label{BaBr3}
 \ee
  Since the conformal transformation is given by $e^{\omega}$,
  the black hole solutions of section \ref{two} for which  $e^{-\omega}=0$, are somehow hidden in action (\ref{BaBr3}).
  From this point of view,  using action (\ref{BaBr3})  to compute the charges of the black hole
  solutions by the method of  \cite{Barnich-Brandt} is questionable.
  Nevertheless, as we show in the following, the central charge obtained by this method is consistent with
  the result of section \ref{three}.
  It is also interesting to note that action (\ref{BaBr3}) can be supplemented by a Gibbons-Hawking term in
  the usual way, which gives the Brown-York stress-tensor used in section \ref{three}.

 In \cite{Barnich-Brandt} it is shown that for Einstein gravity,
  the gravitational conserved
 charge corresponding to an asymptotic Killing vector $\xi$  is given
 by\footnote{For the black hole solutions, $e^{-\omega}=0$, and thus the mass term in Eq.(\ref{BaBr3})
 does not contribute in the energy-momentum tensor. Furthermore  similar to section \ref{three}, we assume that $\delta\omega=0$.}
 \be
 Q_\xi=\lim_{r\to\infty}\int_0^{2\pi}d\theta\,{\tilde k}_\xi^{[tr]}[h,\bar
 g].
 \label{BaBr-charge}
 \ee
 $\bar g$ denote the background metric (the vacuum solution),
  $h_{\mu\nu}$ is the first order deviation of the solution from the background geometry, i.e.
  $g_{\mu\nu}=\bar g_{\mu\nu}+h_{\mu\nu}+{\cal O}(h^2)$ and
 \be
 {\tilde k}_\xi^{[\mu\nu]}[h,\bar g]=\frac{\sqrt{-\bar g}}{2\kappa}\left(\xi_\rho\bar D_\sigma
 H^{\rho\sigma\nu\mu}+\frac{1}{2}H^{\rho\sigma\nu\mu}\partial_\rho\xi_\sigma\right)
 \ee
 in which
 \be
 H^{\mu\alpha\nu\beta}[h,\bar g]={\hat h}^{\alpha\nu}{\bar g}^{\mu\beta}+{\hat h}^{\mu\beta}{\bar
 g}^{\alpha\nu}-(\mu\leftrightarrow\nu)
 \ee
 where
 \be
 {\hat h}_{\mu\nu}=h_{\mu\nu}-\frac{1}{2}\bar g_{\mu\nu}h, \hspace{1cm}h={\bar
 g}^{\mu\nu}h_{\mu\nu}.
 \ee
   For $\xi=(\xi^t,\xi^r,\xi^\phi)$ and
 \be
 h_{\mu\nu}=8GM\,{\rm diag}\left(1,\frac{(1-z)}{2}\left(\frac{r}{\ell}\right)^{2(z-1)},0\right)
 \ee
  one finds
 \be
 Q_\xi=M\,\frac{(1-z)}{2}\frac{\ell}{\cal R}\,\xi^t
 \label{BaBr-Q}
 \ee
 where ${\cal R}$ is the radius of the boundary defined in Eq.\eqref{radius-of-boundary},
 \be
 {\cal R}=\ell\left(\frac{r}{\ell}\right)^{1+z}
 \ee
 Naively, Eq.(\ref{BaBr-Q}) implies that for the black hole solutions studied in section \ref{three}, where $z=1/2$, $Q_\xi=0$, but as we discussed in section \ref{three} (see Eq.\eqref{boundary-coordinates}), the correct value of the charges is given by
 \be
 \tilde Q_\xi=\frac{{\cal R}}{\ell}\,Q_\xi.
 \ee
 Thus $\tilde Q_{\partial_t}=M/4$.

 Furthermore one can compute the central charge for the asymptotic Killing
 vectors. In order to use the formula \cite{Barnich-Brandt}
  \be
 K_{\xi',\xi}=\lim_{r\to\infty}\int_0^{2\pi}d\theta\,{\tilde k}_\xi^{[\mu\nu]}[\delta_{\xi'}g_{\mu\nu},\bar g]
\label{BaBr-K}
 \ee
 to compute the central charge, we consider the vector,
 \be
 \xi=\epsilon\partial_{\cal T}+\lambda\partial_\phi+\alpha
 x\partial_x
 \label{BaBr-Killing}
 \ee
 where
 \be
 \dot \epsilon=\lambda'=-b\alpha,\hspace{1cm}\epsilon'=\dot \lambda.
 \ee
  instead of the asymptotic conformal Killing vectors considered in section
  3. Since
 \be
 \delta_\xi g_{\mu\nu}=\nabla_\mu\xi_\nu+\nabla_\nu\xi_\mu
 \label{BaBr-deltag}
  \ee
 one verifies that similar to section \ref{three}, the correct fall off condition for  $\delta g_{xt}\sim\delta g_{x\phi}\sim
 xg_{xx}$ requires that $b(z+1)<1/2$. Recall that
 \be
 ds^2=x^{2b}(-d{\cal T}^2+d\phi^2)+b^2\ell^2
 x^{2(z+1)b-2}dx^2,\hspace{1cm}x^b=\frac{r}{\ell}.
 \ee
 Of course  the value of the central charge is
 independent of the choice of radial coordinate i.e. it is independent of the value of $b$.
 The difference between the $\delta g_{\mu\nu}$ generated by (\ref{BaBr-Killing}) and the one corresponding to
 the conformal Killing vectors  considered in section \ref{three} is that here,
 \be
 \delta g_{xx}=2\alpha(1+z)g_{xx}.
 \ee
 Recall that the conformal transformation considered in Eq.\eqref{tuned} is tuned to make $\delta g_{xx}=0$.

 Equations (\ref{BaBr-K})-(\ref{BaBr-deltag}) give
 \be
 c=\frac{3{\cal R}}{2G}
 \label{BaBr-c}
 \ee
 which is the value obtained in section \ref{three}.

 \section{Stability of the Schwarzschild solution}\label{ApB}
  The Mart\'{\i}nez-Zanelli solution  is unstable against linear circularly symmetric perturbations \cite{Martinez:1998db}.
  To study the stability of the Schwarzschild solution against linear perturbations, we consider the most general perturbed metric,
 \be
 ds^2=-e^{2U(t,r,\phi)}f(t,r,\phi)dt^2+\frac{dr^2}{f(t,r,\phi)}+2H(t,r,\phi)dtd\phi+r^2d\phi^2
 \ee
 where $f(t,r,\phi)=(1-\frac{2M}{r})+F(t,r,\phi)$. Furthermore we assume that $\psi(t,r,\phi)=\sqrt{\frac{8}{\kappa}}+\xi(t,r,\phi)$.
 Linearizing the Einstein equations with respect to $U(t,r,\phi)$ , $F(t,r,\phi)$, $H(t,r,\phi)$ and $\xi(t,r,\phi)$ one obtains,
 \bea
 \label{Gtphi}
 0&=&\frac{\partial^2}{\partial t\partial\phi}\xi,\\
 \label{Gtt}
 0&=&r^2(r-2M)\frac{\partial^2}{\partial r^2}\xi+r\frac{\partial^2}{\partial \phi^2}\xi+
 r(r-M)\frac{\partial}{\partial r}\xi+M\xi,\\
 \label{Grr}
 0&=&r(r-2M)\frac{\partial^2}{\partial \phi^2}\xi-r^4\frac{\partial^2}{\partial t^2}\xi+(r-2M)\left(r(r-M)\frac{\partial}{\partial r}\xi+M\xi\right),\\
 \label{Gphiphi}
 0&=&r^2(r-2M)^2\frac{\partial^2}{\partial r^2}\xi-r^4\frac{\partial^2}{\partial t^2}\xi+2M(r-2M)\left(-\xi+r\frac{\partial}{\partial r}\xi\right).
 \eea
  Eqs.\eqref{Gtt}-\eqref{Gphiphi} give,
 \bea
 \label{Ct}
  \frac{\partial^2}{\partial t^2}\xi&=&\frac{M(r-2M)}{r^4}\left(-\xi+r\frac{\partial}{\partial r}\xi\right)\\
  \label{Cr}
 \frac{\partial^2}{\partial r^2}\xi&=&-\frac{M}{r^2(r-2M)}\left(-\xi+r\frac{\partial}{\partial r}\xi\right),\\
 \label{Cphi}
  \frac{\partial^2}{\partial \phi^2}\xi&=&-\left(\frac{2M}{r}\xi+(r-2M)\frac{\partial}{\partial r}\xi\right).
 \eea
  The only possible solution of these equations is,
  \be
  \xi(t,r,\phi)=r\Phi(\phi),\hspace{1cm}\frac{\partial^2}{\partial \phi^2}\Phi(\phi)=-\Phi(\phi).
  \ee
 In this case, the  equation of motion of $\psi$ simplifies as
  \bea
 0&=&r(r-2M)^3\frac{\partial^2U}{\partial r^2}+\frac{r^2}{2}(r-2M)^2\frac{\partial^2F}{\partial r^2}+(r-2M)^2\frac{\partial^2U}{\partial \phi^2}
 \nn\\&+&
  r(r-2M)\frac{\partial^2H}{\partial t\partial\phi}+\frac{r^4}{2}\frac{\partial^2F}{\partial t^2}+(r-2M)^2
  \left((r+M)\frac{\partial U}{\partial r}+r\frac{\partial F}{\partial r}\right).
  \eea
  which also gives $R=0$. Near the horizon this equation simplifies as follows,
  \bea
 0&=& Mx^2\left(2x\frac{\partial^2U}{\partial x^2}+3\frac{\partial U}{\partial x}+\frac{1}{M}\frac{\partial^2U}{\partial \phi^2}\right)\nn\\
 &+& 2M^2\left(x^2\frac{\partial^2F}{\partial x^2}+4M^2\frac{\partial^2F}{\partial t^2}\right)+2Mx\frac{\partial^2H}{\partial t\partial\phi},
 \label{eq}
  \eea
  where $x\equiv r-2M$. Assume that $H(t,r,\phi)=0$,  $F(t,r,\phi)=\epsilon(t,\phi)x^p$, and $U(t,r,\phi)=M^{p-q}u(t,\phi)x^q$, where $p,q\ge0$.
  If $q\neq p-1$
  then the first term in Eq.\eqref{eq} implies that $u(t,\phi)=0$, and consequently,
  \be
  \frac{\partial^2\epsilon(t,\phi)}{\partial t^2}=-\frac{p(p-1)}{4M^2}\epsilon(t,\phi).
  \ee
  Thus, modes corresponding to $0<p<1$ which radial slope diverges on the event horizon, grow with time. This shows that the Schwarzschild solution is
  unstable against such perturbations. It is useful to note that this condition is also sufficient to prevent curvature
  singularities at the event horizon.

  For $q=p-1$ one obtains,
 \be
  Mx^2\left(2x\frac{\partial^2U}{\partial x^2}+3\frac{\partial U}{\partial x}\right)+
  2M^2\left(x^2\frac{\partial^2F}{\partial x^2}+4M^2\frac{\partial^2F}{\partial t^2}\right)=0,
 \label{eq1}
  \ee
 which gives
 \be
 4M^2  \frac{\partial^2\epsilon(t,\phi)}{\partial t^2}=-\left[q(q+1)\epsilon(t,\phi)+q(q+\frac{1}{2})u(t,\phi)\right].
  \ee
 In principle, $u(t,\phi)=\alpha\epsilon(t,\phi)$. Thus for $\alpha<-(1+\frac{1}{2q+1})$, these modes also grow with time.


\end{document}